\pdfoutput=1
\documentclass[pdflatex,sn-mathphys-num]{sn-jnl}

\usepackage{graphicx}%
\usepackage{multirow}%
\usepackage{amsmath,amssymb,amsfonts}%
\usepackage{amsthm}%
\usepackage{mathrsfs}%
\usepackage[title]{appendix}%
\usepackage{xcolor}%
\usepackage{textcomp}%
\usepackage{manyfoot}%
\usepackage{booktabs}%
\usepackage{algorithm}%
\usepackage{algorithmicx}%
\usepackage{algpseudocode}%
\usepackage{listings}%

\theoremstyle{thmstyleone}%
\theoremstyle{thmstyletwo}%

\theoremstyle{thmstylethree}%

\begin{document}

\title[Agent]{Generative Organizational Behavior Simulation using Large Language Model based Autonomous Agents: A Holacracy Perspective}


\author[1]{\fnm{Chen} \sur{Zhu}}\email{zc3930155@gmail.com}
\equalcont{These authors contributed equally to this work.}

\author[2]{\fnm{Yihang} \sur{Cheng}}\email{chengyihang544@gmail.com}
\equalcont{These authors contributed equally to this work.}

\author[2]{\fnm{Jingshuai} \sur{Zhang}}\email{zhangjingshuai0@gmail.com}
\equalcont{These authors contributed equally to this work.}

\author[3]{\fnm{Yusheng} \sur{Qiu}}\email{qiuyusheng@csu.edu.cn}

\author[4]{\fnm{Sitao} \sur{Xia}}\email{stxia@tju.edu.cn}

\author*[2,5]{\fnm{Hengshu} \sur{Zhu}}\email{zhuhengshu@gmail.com}

\affil[1]{\orgdiv{University of Science and Technology of China}, \orgname{School of Management}}

\affil[2]{\orgdiv{Career Science Lab}, \orgname{BOSS Zhipin}} 


\affil[3]{\orgdiv{School of Computing}, \orgname{Central South University}}

\affil[4]{\orgdiv{College of Management and Economics}, \orgname{Tianjin University}}

\affil[5]{\orgdiv{Computer Network Information Center}, \orgname{Chinese Academy of Sciences}}


\abstract{\abstract{abstract}
In this paper, we present the technical details and periodic findings of our project, CareerAgent, which aims to build a generative simulation framework for a Holacracy organization using Large Language Model-based Autonomous Agents. Specifically, the simulation framework includes three phases: construction, execution, and evaluation, and it incorporates basic characteristics of individuals, organizations, tasks, and meetings. Through our simulation, we obtained several interesting findings. At the organizational level, an increase in the average values of management competence and functional competence can reduce overall members' stress levels, but it negatively impacts deeper organizational performance measures such as average task completion. At the individual level, both competences can improve members' work performance. From the analysis of social networks, we found that highly competent members selectively participate in certain tasks and take on more responsibilities. Over time, small sub-communities form around these highly competent members within the holacracy. These findings contribute theoretically to the study of organizational science and provide practical insights for managers to understand the organization dynamics.

}




\maketitle

\section{Introduction}\label{sec1}

\begin{figure}
    \centering
    \includegraphics[width=1.0\linewidth]{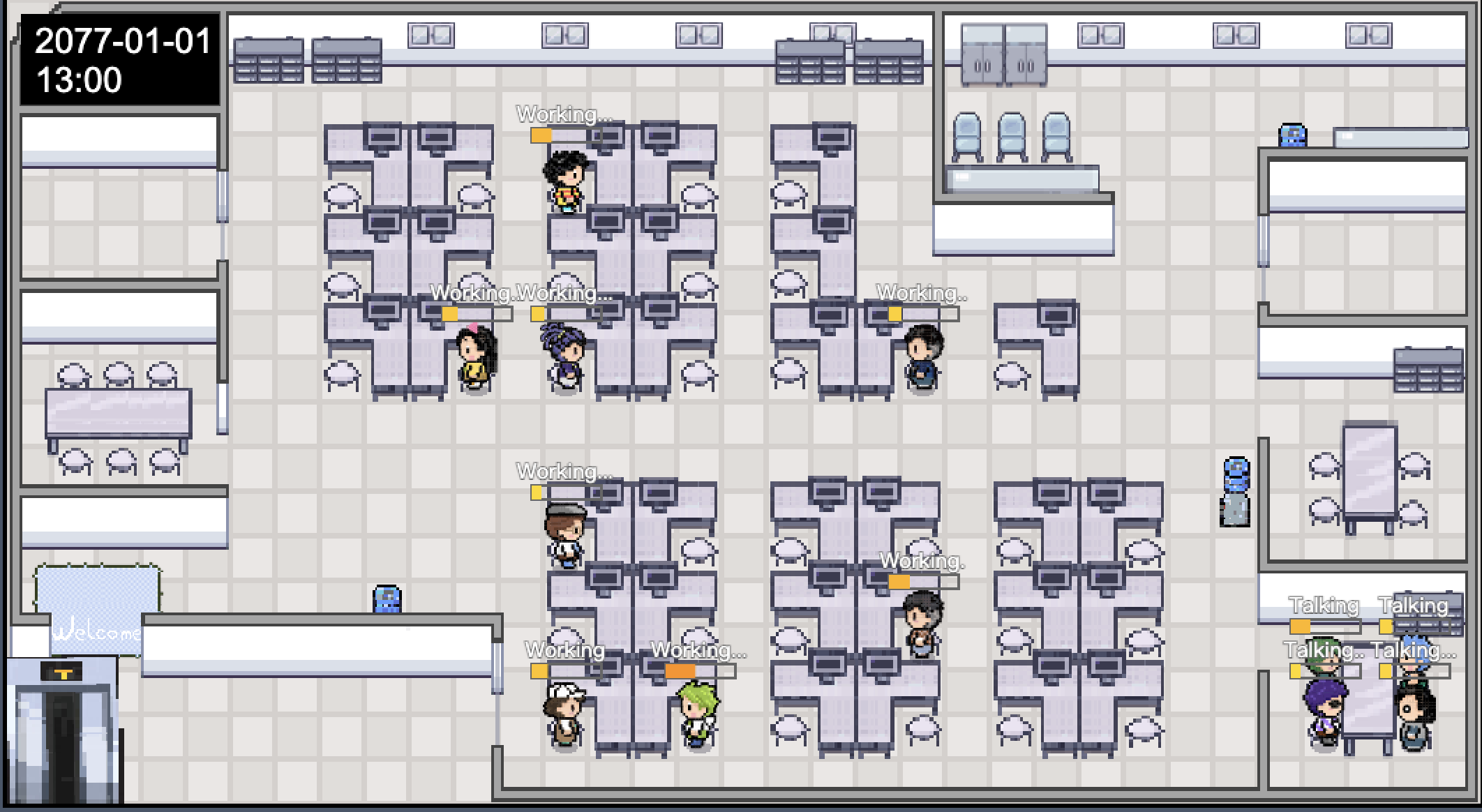}
    \caption{Generative agents in an organization}
    \label{fig:Screen_example}
\end{figure}

Holacracy is an innovative management model proposed by Brian Robertson, the founder of the software company. It is a democratic and open organizational structure with shared governance for all, aiming at the decentralized management of an organization by breaking the authoritarianism of the leadership through the assumption of work by roles\cite{van2014holacracy}. Such a management model is better to give employees the freedom to be more creative; however, at the same time, it also creates conflicts between roles and teams, resulting in many organizational practices ending in failure \cite{schell2022change}. 
Although some static influence mechanisms have been explored in the past \cite{ackermann2021mercedes, oe2023innovative}, the dynamic operation of the system, like autority delegation, is not well understood.
In this paper, based on the simulation capacity of Large Language Model (LLM) \cite{xi2023rise,mandi2023roco}, we built \textit{CareerAgent}, an organizational behavior simulation framework based on LLM Agents, as shown in Figure \ref{fig:Screen_example}, to simulate the operation of organizations under the holacracy framework, and found some interesting phenomena.

One of the characteristics of the holacracy is that the leaders delegate their authority to the employees at the lower level. This act of authorization puts the work or project in the hands of the employees and gives them full autonomy in decision-making, which leads to pressure increase for the employees\cite{porcelli2017stress}. Work stress is characterized by two components, psychological workload and decision-making volume, and the interaction of workload and decision-making volume produces psychological stress \cite{schnall1994job,wall1996demands}. Under such a characterization, employees' managerial decision-making competence and professional competence to deal with work will have an impact on their perception of stress. At the same time, different employee competence characteristics will also affect the experience of cooperation between individuals, which in turn will affect the division of the circle \cite{bernstein2016beyond}.

Therefore, we conducted a simulation experiment based on the LLM-based Agent approach for eight weeks, with weekly rounds of task issuance and billing, and four tasks per round. The study worked through eight weeks of simulations and found results at both the organizational and individual levels. 
At the organizational level, average competence has a positive effect on work completion and a negative effect on stress. At the circle level, the higher the average competence, the lower the average number of circles per employee and the lower the average number of people in each circle. The higher the average competence, the smaller the average degree of employee communication network formation in the organization. At the individual level competence has a positive effect on work completion and a negative effect on stress. The higher the competence the smaller the number of circles in which they are located and the greater the average workload.

\section{The process and key concepts of holacracy}\label{sec4}
The holacracy is a management framework with a set of well-defined rules that detail the "rules of engagement" for each job in the organization, including the organizational structure, decision-making, and the way power is distributed. One of the main features of the holacracy is "democracy", which means that the leader needs to decentralize his/her authority, eliminating the traditional CEO, and decentralizing management responsibilities to different roles to achieve "autonomy" for employees. Specifically, the "constitution" of a holacracy can be divided into five parts: organizational structure, rules of cooperation, tactical meetings, distribution of authority, and governance processes.

\vspace{0.1cm}
\noindent\textbf{(1) Organizational structure}
\vspace{0.1cm}

A \textbf{“Role”} is an organizational construct that a person can fill and then energize on behalf of the Organization. Whoever fills a Role is a “Role Lead” for that Role. A “Role” is responsible for regularly considering how to enact your Role’s Purpose and each Accountability, by defining: “Next-Actions”, which are useful actions that you could take immediately, at least in the absence of competing priorities; “Projects”, which are specific outcomes that would be useful to work towards, at least in the absence of competing priorities.

A\textbf{ “Circle”} is a container for organizing Roles and Policies around a common Purpose. The Roles and Policies within a Circle make up its acting “Governance”. The inside of every Role is a Circle. This Circle can hold its own Roles and Policies to break down and organize its work. This does not apply to the Roles defined in this Constitution, which may not be further broken down. A Role’s internal Circle is considered a “Sub-Circle” of the broader Circle that holds the Role, while that broader Circle is its “Super-Circle”.

Any Circle may appoint someone as the Circle’s\textbf{ “Facilitator”}. The selected Facilitator fills a “Facilitator Role” in the Circle, with a Purpose of “Circle governance and operational practices aligned with the Constitution”.

Any Circle may appoint someone as the Circle’s \textbf{“Secretary”}. The selected Secretary fills a “Secretary Role” in the Circle, with a Purpose of “Stabilize the Circle’s constitutionally-required records and meetings”. A Circle may add Accountabilities or Domains to its own Facilitator or Secretary Role, as well as amend or remove those additions. No Circle may amend or remove the Purpose of either Role, nor any Accountabilities or Domains placed on those Roles by this Constitution.

\vspace{0.1cm}
\noindent\textbf{(2) Rules of cooperation}
\vspace{0.1cm}

The rules of cooperation in the holacracy define the basic duties that all partners expect from each other. It includes duty of transparency (e.g., upon request, all roles must share what they are working on); duty of processing (e.g., each role must respond to a request to accept new work and accept it unless they decide not to, in which case they must explain why), duty of prioritization (e.g., must be aligned with the circle's strategy, prioritize securing internal communications, etc.), and relational agreements, which are agreements about how you will relate together while working in the organization, or about how you will fulfill the general functions as Partners of the Organization. 

\vspace{0.1cm}
\noindent\textbf{(3) Tactical meetings}
\vspace{0.1cm}

Any Partner may convene a\textbf{ “Tactical Meeting”} to assist Partners in engaging each other in their responsibilities and duties. In addition, the Secretary of each Circle is accountable for scheduling regular Tactical Meetings for the Circle. The Facilitator of a Circle is accountable for facilitating the Circle’s regular Tactical Meetings, and its Secretary is accountable for capturing and publishing Tactical Meeting outputs. For Tactical Meetings convened by someone other than a Circle’s Secretary, the Partner convening a Tactical Meeting must facilitate it and capture its outputs, or appoint another volunteer or appropriate Role to do so.

\vspace{0.1cm}
\noindent\textbf{(4) Distributed authority}
\vspace{0.1cm}

As a Role Lead, you have the authority to take any action or make any decision to enact your Role’s Purpose or Accountabilities, as long as you don’t break a rule defined in this Constitution. When prioritizing and choosing among your potential actions, you may use your own reasonable judgment of the relative value to the Organization of each.

\vspace{0.1cm}
\noindent\textbf{(5) Governance process}
\vspace{0.1cm}

Changing a Circle’s Governance requires using the\textbf{ “Governance Process”} defined herein. The holacracy governance process provides for decentralized governance processes within each team. Anyone can suggest improvements to the structure of the organization within their scope of work, and it allows anyone in the organization (not just "leaders") to suggest improvements to the organizational structure of roles and circles. Such a rule fundamentally shifts the center of power to the governance process, making the organization truly "self-governing" and "purpose-driven".

\section{LLM-based simulation platform for holacracy}\label{sec5}

In this section, we introduce our LLM-based simulation framework for holacracy. A Large Language Model (LLM) is a deep learning-based natural language processing model capable of producing high-quality text. LLM uses a large amount of training data to learn and has the ability to understand and generate language. Its text generation capability can be simplified as the following equation:
\begin{equation}
    \text{Text}_{\text{output}} = \text{LLM}(\text{Text}_{\text{input}})
    \label{eq:framework_llm}
\end{equation}
where, $\text{Text}_{\text{output}}$ is the generated text, $(\text{Text}_{\text{input}})$ is the input text.

\begin{figure}
    \centering
    \includegraphics[width=1.0\linewidth]{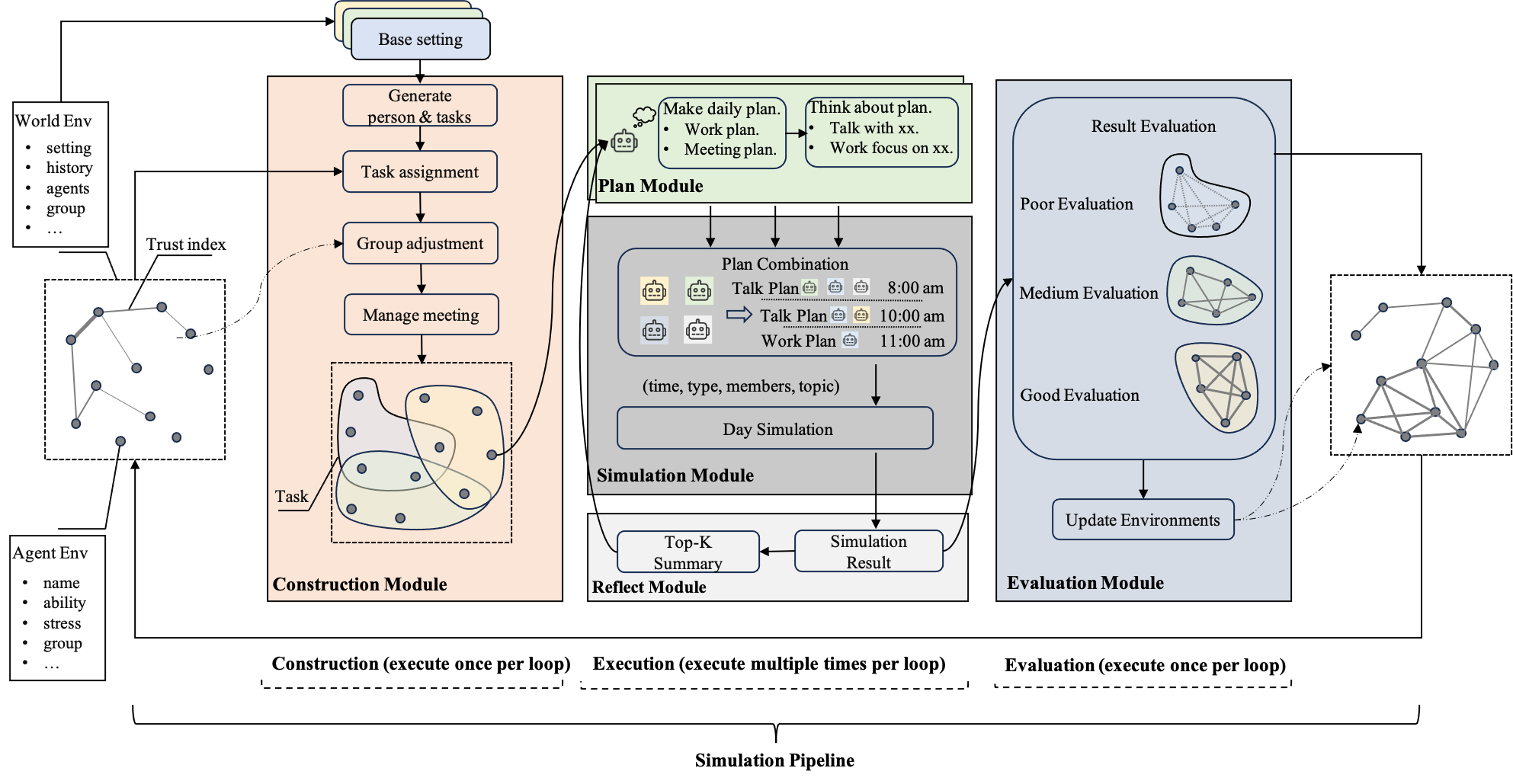}
    \caption{Simulation framework overview}
    \label{fig:Process_example}
\end{figure}

LLM has strong simulation ability of role. In practical application, it can simulate various roles and generate corresponding dialogues and behaviors according to the set tasks and scenes. Using the simulation power of the LLM, we divide the holacracy simulation into three phases: (1) Construction phase: Employees form specific working relationships/connections according to the rules of the organization, which is the beginning part of the pipeline and is performed once in a cycle (initialization); (2) Execution phase: Employees perform corresponding jobs/tasks according to specific working relationships, which is the main part of the whole pipeline and will be executed many times in a cycle; (3) Evaluation phase: Evaluation and feedback of current members and their relationships based on their work completion, which further influences the next cycle. The specific framework is displayed in Figure \ref{fig:Process_example}, which has a context background, including members' characteristics, world environment, and members' initial relationships. Next, we will introduce the specific information about different phases.

\subsection{Construction phase}
The construction phase is mainly to enter the context of the organization, members, and world for the framework, while carrying out the initial task generation and assignment.

\indent\textbf{World environment setting}
includes the target, industry, tasks of the organization, this parameters could be represented by a set $E$.

\indent\textbf{Members generation}
is set according to the world environment. There are $N$ members, each of which includes position, skills, competences, experiences and so on, these characteristics could be represented by $\mathbf{e}_i$:
\begin{equation}
    \mathbf{e}_i = [e_{i1}, e_{i2}, \ldots, e_{im}]
\end{equation}
where, $e_{ij}$ is the $j_{th}$ characteristic of the member $i$, there are $m$ of the member $i$. The member generation process could be formulated as follows:
\begin{equation}
    \{e_i\}_{i=1}^N = \text{LLM}(\mathbf{E}, \text{Generate } N \text{ persons})
\end{equation}
where $\text{LLM}$ is the generation function based on LLM.

\indent\textbf{Task generation}
is generated according to the world environment and the number of members $N$, the number of task is $M$ and the workload of each task is $WT$. Workload refers to the required working hours. These parameters satisfy the following relationship:
 \begin{equation}
     N * 8 * 5 = M * WT
 \end{equation}
where there are five workdays and each of workday has eight working hours. The contents of these tasks are generated by LLM:
\begin{equation}
    \{T_i\}_{i=1}^M = \text{LLM}(\mathbf{E}, \text{Generate } M \text{ tasks})
\end{equation}
where $T_i$ is the $i_{th}$ task. 

\indent\textbf{Task allocation}
is conducted by LLM according to the number of members required for tasks, members' characteristics, the number of tasks assigned by the current members and the history of cooperation. As for task $T_j$, the range of the number of required members is $[N_{j,\min}, N_{j,\max}]$, the number of tasks assigned to the current employee is $Q_i$, the trust formed through cooperation history could be expressed as a matrix:
\begin{equation}
    \mathbf{H} = [h_{ij}]
\end{equation}
where $h_{ij}$ is the trust value between member $i$ and member $j$. Then, these parameters could be input into the LLM, the members allocation in task $T_j$ could be represented by a vector:
\begin{equation}
    \mathbf{a}_j = \text{LLM}(T_j, [N_{j,\min}, N_{j,\max}], \{e_i\}_{i=1}^N, \mathbf{H}, \mathbf{Q})
\end{equation}
where $\mathbf{Q} = [Q_i]$ is the assignments situation for all members, the length of $\mathbf{a}_j$ is $N_j$, representing the members assigned to the task $T_j$.


\indent\textbf{Circle adjustment}
is conducted by members' voting within the circle, which is controlled by the LLM. Then, the new members assignments vector has been obtained:
\begin{equation}
    \mathbf{a}_j' = \text{LLM}(T_j, \mathbf{H}, \mathbf{a}_j)
\end{equation}

\indent\textbf{Governance meeting}
is used to assign roles of the task $T_j$ to members, the role could be represented by a vector, i.e., $\mathbf{r}_j = [r_{ij}]$, where $r_{ij}$ is the role of the task $T_j$ assigned to member $i$. Through the LLM, $\mathbf{r}_j$ could be represented by:
\begin{equation}
    \mathbf{r}_j = \text{LLM}( T_j, \mathbf{a}_j', \{e_i\}_{i=1}^N)
\end{equation}

Through the above steps, the framework completed the task assignment phase, assigning the right members to each task and clarifying their roles, providing an organizational foundation for the subsequent work.



\subsection{Execution phase}

After each member has a clear task and the corresponding circle role responsibilities, the multi-round implementation stage is conducted.

\indent\textbf{Work planning}
is conducted by LLM-based member according to task and role responsibilities. Work plan $\mathbf{P}_i$ of member $i$ includes working hours, task and partners, which could be represented by:
\begin{equation}
    \mathbf{P}_i = \{ (t_{ij}, T_j, p_{ij}) \}
\end{equation}
where $t_{ij}$ is the working hour, $p_{ij}$ is partners. $\mathbf{P}_i$ is generated by LLM:
\begin{equation}
    \mathbf{P}_i = \text{LLM}( T_j, \mathbf{a}_j', C_j, WR_i)
\end{equation}
where $C_j$ is the current completion of the task $j$, $WR_i$ is the work records in this cycle of the member $i$.



\indent\textbf{Plan consolidation}
is set to ensure that no duplicate tasks occur, resulting in duplicate workload. For example, if member $i$ wants to chat with member $j$ and member $i$ wants to chat with member $j$, the the framework will combine the plans of $i$ and $j$ into one plan $\mathbf{P}_{ij}$. $\mathbf{P}_{ij}$ includes all plans of the member $i$ and member $j$. Then the framework will combine all individual plans to form a unified world plan $\mathbf{WP}$.
%

\indent\textbf{Plan execution}
sorted according to the fixed world plan $\mathbf{WP}$, after which we can get the simulation results of the day, including the work of the members on different tasks, chat history, meeting discussion and so on

\indent\textbf{Information summary}
is used to form the member's memory $\mathbf{M}_i$, this process is conducted by the LLM. Final individual memory includes the work records and the history of cooperation. These memories could be input into the next step to influence the sequential members' behaviors.


\subsection{Evaluation phase}
After completing a week's work simulation through multiple execution phases, the following steps are taken to evaluate task completion and member's performance.

\indent\textbf{Task completion evaluation}
is conducted according to the individual memories. In this paper, $C_j$ is the task completion level, calculated by the ratio of the total working hours spent by the members participating in the task to the total working hours required by task $T_j$.


\indent\textbf{Member's evaluation}
is conducted according to the completion level of all tasks assigned to the member $i$:
\begin{equation}
    D_i = \frac{1}{k} \sum_{j=1}^{k} C_{ij}
\end{equation}
where $k$ is the number of tasks that member $i$ participates in, and $C_{ij}$ is the performance of member $i$ participates in task $T_j$.


\indent\textbf{Framework information update}
is set to ensure that the information stored in the framework is kept up to date, and that this information will have a continuous impact on the behavior of members in the next stage. Specifically, new work records ${WR}'$ could be generated by all work records in this cycle.
\begin{equation}
    {WR}' = \text{LLMSummary}(WR)
\end{equation}
Besides, the level of members' trust is related to the performance of cooperative members. Poor evaluation results will weaken the level of trust among members, whereas good results will strengthen the level of trust, therefore, new trust matrix $\mathbf{H}'$ could be generated by current trust matrix $\mathbf{H}$ and current individual evaluation results $\mathbf{D}$. 








\section{Experimental design}\label{sec2}
\subsection{Research issues}
Holacracy seems to be a more democratic and free form of organization that can maximize the advantages, but the practice in the organization also reveals a lot of problems, which can be seen that there are two sides of the holacracy, which makes its application have a lot of unknown risks\cite{bernstein2016beyond}\cite{schell2022change}. At the same time, the attempt in practice will spend a lot of manpower and material resources, there is a long test cycle, high cost and many other inconveniences. Traditional approaches to theory development are limited in their ability to analyze multiple interdependent processes operating simultaneously\cite{harrison2007simulation}. As technology develops, LLM becomes more and more mature, and the research on AI for simulating the real world continues to progress, making it possible to use Agents to simulate organizational practices\cite{gurcan2024llm}. Therefore, we try to explore the effect of employee competence on work pressure, completion and circle division in the organization under the management mode of holacracy by using a large model-based Agent simulation method. Due to the non-role differences in individual attributes, the individual employee attributes in the experiment are divided by individuals rather than roles, but different roles involve different work and decision-making practices.

One of the characteristics of holacracy is that leaders delegate their rights to the lower level employees. Such an act of authorization puts the work or project under the responsibility of the employees and gives them full autonomy in decision-making, which is both an opportunity and a pressure for the employees\cite{krasulja2016holacracy}. On the one hand, authorizing the leader can increase the positive perception of the employees and motivate them to work actively; but on the other hand, successfully completing the work authorized by the leader is also a kind of pressure on the employees.

Job stress is characterized by two components, the psychological job demand and decision-making dimensions, and the interaction between job demand and decision-making latitude produces psychological stress\cite{schnall1994job}. It has been found that perceived workload predicts job satisfaction and absenteeism, and that the interaction between the objective demand index (derived from job analysis) and perceived control predicts sick leave and tardiness\cite{wall1996demands}. With such characteristics, employees' managerial decision-making ability and professional competence in handling their work will have an impact on their perceived stress.

When the employee is competent, the empowered leader, by sharing power with the employee and allowing the employee to exercise autonomy, encourages the employee to experience positive feedback from the organization or the team, and to view the work of the organization as his or her own, and to return positively to the organization's recognition and respect, thus achieving a good level of performance. On the contrary, when the employee is weak, the empowered leader can become a burden to the employee because the employee believes that he or she does not have the ability to complete the work independently and autonomously, and in this situation, the employee will not produce a high level of performance, but rather, it will also increase the pressure.

From the above discussion, it is evident that employees' competence is a critical determining factor in the success of an organization. Therefore, in this study, we explore the impact of employees' competence at both the organizational and individual levels. At the organizational level, we analyze how the average competence level within the organization affects the state of stress, work completion, and circle formation within the organization. At the individual level, we examine how different levels of individual competence influence personal stress, work completion, workload, and the formation of social circles.


\subsection{Parameter design and variable measurements}

The simulation experiment design process includes two parts: the workday cycle and the settlement session. The workday includes processes such as task distribution, governance meeting, tactical meeting, work, discussion, etc.; the settlement link mainly involves the summary of changes in individual state, acceptance of the degree of completion of the work, memory organization and other modules. The simulation experiment design includes two parts: individual and organization. The organizational framework is based on the holacracy to design the corresponding task allocation, meeting, work and other related processes; the individual plate involves three parts: inherent labels, variable attributes, and actions. On the basis of simplifying the real world, we try to simulate the real organizational work state from multiple dimensions.

\subsubsection{Individual}
The individual segment contains three aspects of the individual's inherent attributes, variable states, and behaviors, and is represented in the data information by two parts: personal information and personal records.

\indent\textbf{Inherent attributes:}
Personality descriptions describing character as well as life habits, such as mild, lively, calm, lazy, etc., and life habits expressed as whether or not they take lunch breaks/like to exercise, etc., affect individual performance.
Personal skills includes personal research direction description, management competence and functional competence. Personal skills will affect the individual's state and decision-making. Among them, management competence refers to the ability of individual management, including self-management, decision-making ability; functional ability refers to the degree of individual competence in the current position, including job-specific specialized theoretical needs of knowledge reserves.

\indent\textbf{Variable state:}
Work pressure, mainly reflects the individual's demand for work and decision-making, i.e., an increase in the amount of work and decision-making will increase the value of individual's pressure. At the same time, different individuals have different sensitivities to stress according to their competence, i.e., the value of pressure increases differently when facing the same amount of work or decision-making.
Work completion, i.e. the difference between actual work results and expectations. The level of work completion affects the willingness of individuals to cooperate with each other, which in turn affects subsequent circle divisions.

~~\noindent\textbf{Individual actions:}
Governance meeting is to identify the various roles currently undertaken by individuals and the circle situations in which they are located;
Tactical meeting is to solve problems encountered in the work;
Work behaviors include behaviors such as making daily plans, reflecting and remembering, and simulating real-world work by accumulating work hours.

\subsubsection{Organization}
The organization board contains aspects such as base settings, project lists, and circle divisions.

\indent\textbf{Base settings:}
Organizational background setting is used to record the basic information of the organization. The management mode is holacracy, which is mainly reflected in the freedom of task distribution, high transparency of information, and the role/circle structure. The organizational goal is the goal of the fundamental circle, which is the behavioral quasi-measurement all the staffs have to comply with as a priority, and the goal of the current stage is to complete the list of all the projects.

\indent\textbf{Project lists:}
The project list contains information such as work hours required for completion, deadline dates, and task profiles. Tasks are issued, accepted and settled at certain time intervals. Here is an example,

``Task": ``Research and implement a new artificial intelligence algorithm".

``Description": ``Research and implement an AI algorithm that can be effectively applied to recruitment scenarios, capable of intelligently assessing and providing recommendations based on a candidate's resume and interview performance"

``Work Hours": ``100"

\indent\textbf{Circle divisions:}
Record the division of circles in each round of the project, including the division of labor among agents, coordinators, secretaries, and other members of the circle. The division of circles was judged according to the average number of circles and the average workload, with a larger average number of circles indicating that individuals chose to join more circles on average, i.e., a larger number of tasks, and a higher average workload indicating that there were fewer people in the circle, and that on average, each person was assigned more work.

\subsection{Experimental implementation}
The experiments were conducted based on different levels of organizational average capabilities, which include two categories for management competence (high and low) and two categories for functional competence (high and low). Additionally, variations in competence variance were considered, with three categories each for management competence (high, medium, and low) and functional competence (high, medium, and low). This resulted in a total of 36 (2*2*3*3) experimental groups. Each experiment lasted for eight weeks. During each week of the experiment, tasks were assigned and selected, with each round consisting of four tasks.

\section{Results}\label{sec3}
To explore the research issues in this paper, we conducted two aspects of analysis. On the one hand, we use the econometric model to evaluate the theoretical mechanism of the Holacracy. On the other hand, we analyzed the social network dynamics during the simulation.

\subsection{Empirical analyses}\label{empirical}
Firstly, we specified the following econometric model (Equation \ref{eq:reg_organizaiton}) to estimate the influence of degree of two different competences in organization on the work completion level, average stress degree and average circle number:
\begin{equation}
Y_i = \beta_0 + \beta_1 Management\_competence_i + \beta_2 Functional\_competence_i + \varepsilon
\label{eq:reg_organizaiton}
\end{equation}
\noindent where $Y_i$ is the situation level for organization $ i $, including three dependent variables, which are work completion level, average stress degree and average circle number at the end of the eighth week.  
Since these three variables are the continuous variables, we used ordinary least squares regression (OLS) to estimate $Y_i$, which were measured using ordinal scales. 
At the same time, $Management\_competence$ and $Functional\_competence$ represent the competence level for organization $ i $, including two variables correspondingly, which are the average value and variations value measured by standard deviation at the end of the eighth week. $\beta_1$ and $\beta_2$ are thus the coefficients of interest at the organizational level.

The results of organizational level are shown in Table \ref{tab:main_organization}. According to the results, two kinds of competence mean all have a negative effect on the average stress degree ($Management\_competence: \beta_1=-8.417, p<0.01; Functional\_competence: \beta_2=-8.099, p<0.01$), i.e., the higher the average competence of the organization, the lower the stress felt by the members of the organization. Besides, two kinds of competence mean all have a negative effect on work completion in the organization as a whole ($Management\_competence: \beta_1=-0.080, p<0.05; Functional\_competence: \beta_2=-0.084, p<0.05$), i.e., when the average competence of the organization is stronger, however, the task completion in the organization is worse. This result suggests that a relatively high mean competence of members in an organization does not necessarily lead to efficient work completion, but will cause competition or task allocation problems in some organizations, resulting in a decline in the level of work completion. In terms of circle segmentation and competence variations, there is no significant effect among these relationships. Overall, the average competence of the members in the organization does not necessarily completely determine the organizational performance, but more can determine the direct performance, i.e. members' stress, and the organization's competence, i.e., the work completion, may also require other factors \cite{argote2011organizational,sturm2021coordinating}.

\begin{table}[h!]
    \centering
    \caption{Regression at the 8th week at organization level}
    \begin{tabular}{lcccc}
     \hline
       & Average stress & Work completion level & Average circle number \\
     \toprule
     Management competence mean & $-8.417^{***}$ & $-0.080^{**}$ & $0.020$ \\
     Functional competence mean & $-8.099^{***}$ & $-0.084^{**}$ & $0.003$ \\
     Management competence std & $1.839$ & $0.047$ & $-0.008$ \\
     Functional competence std & $0.621$ & $-0.017$ & $0.000$ \\
     N & $36$ & $36$ & $36$  \\
     $R^{2}$ & $0.913$ & $0.330$ & $0.036$ \\
     \bottomrule
    \end{tabular}
        \begin{tablenotes}
            \item Note:$^{*}p\textless0.1$,$^{**}p\textless0.05$,$^{***}p\textless0.001$
        \end{tablenotes}
    \label{tab:main_organization}
\end{table}

Secondly, we specified the following econometric model (Equation \ref{eq:reg_individual}) to estimate the influence of degree of two different competences of the members on the corresponding work completion level, average stress degree and average circle number. Furthermore, at the individual level, in particular, we added the average workload to measure the performance of different members:
\begin{equation}
Y_i = \beta_0 + \beta_1 Management\_competence_i + \beta_2 Functional\_competence_i + \varepsilon
\label{eq:reg_individual}
\end{equation}
\noindent where $Y_i$ is the situation level for member $ i $, including four dependent variables, which are stress degree, work completion level, average circle number and average workload at the end of the eight week.  
Also, these variables are the continuous variables, we used ordinary least squares regression (OLS) to estimate $Y_i$, which were measured using ordinal scales. 
At the same time, $Management\_competence$ and $Functional\_competence$ represent the competence level for organization $ i $, as for individual, which are only the mean values. $\beta_1$ and $\beta_2$ are thus the coefficients of interest at the organizational level.

The main results of individual level are shown in Table \ref{tab:main_individual}. Accordingly, two kinds of competence all have a negative effect on the stress degree ($Management\_competence: \beta_1=-9.259, p<0.01; Functional\_competence: \beta_2=-6.479, p<0.01$). Besides, two kinds of competence all have a positive effect on individual work completion ($Management\_competence: \beta_1=0.185, p<0.01; Functional\_competence: \beta_2=0.550, p<0.01$). Interestingly, we could find some results different with the organizational level. Both competencies positively influence the average circle number ($Management\_competence: \beta_1=0.177, p<0.01; Functional\_competence: \beta_2=0.558, p<0.01$) and also have a positive impact on average workload ($Management\_competence: \beta_1=0.016, p<0.01; Functional\_competence: \beta_2=0.048, p<0.01$), suggesting that individuals with stronger competencies take on more tasks while fewer people complete each task simultaneously. Furthermore, the results show that high-level members tend to selectively build appropriate circle sizes rather than blindly act collectively.
Overall, within the holacracy, it is evident that competence motivates individuals to voluntarily assume more responsibilities. Additionally, individual results can partially elucidate the results at organizational level. When the individual competence of the organization is high, that is, the organization has a high average competence, too many tasks for each member will lead to an increase in coordination cost and an insurmountable management competence, which will lead to a decline in the level of work completion level. Furthermore, in order to better observe the interaction between the two competences, we observed the impact of the mean value and difference of competences on different outcome variables, and the specific results are shown in Table \ref{tab:main_individial_statistics}. According to the results, we can see that the influence effect of the mean is same as the main results, however, the difference has no significant impact. 

 
\begin{table}[h!]
    \centering
    \caption{Regression at the 8th week at individual level}
    \begin{tabular}{lp{1.3cm} cp{1.3cm} cp{1.3cm} cp{1.3cm} cp{1.3cm}}
     \hline
       & Stress & Work completion level & Average circle number & Average workload \\
     \toprule
     Management competence & $-9.259^{***}$ & $0.185^{***}$ & $0.177^{***}$ & $0.016^{***}$ \\
     Functional competence & $-6.479^{***}$ & $0.550^{***}$ & $0.558^{***}$ & $0.048^{***}$ \\
     N & $720$ & $720$ & $720$ & $720$ \\
     $R^{2}$ & $0.095$ & $0.121$ & $0.134$ & $0.127$ \\
     \bottomrule
    \end{tabular}
     \begin{tablenotes}
            \item Note:$^{*}p\textless0.1$,$^{**}p\textless0.05$,$^{***}p\textless0.001$
        \end{tablenotes}
    \label{tab:main_individual}
\end{table}

\begin{table}[h!]
    \centering
    \caption{Regression at the 8th week at individual level of competence statistics}
    \begin{tabular}{lp{1.3cm} cp{1.3cm} cp{1.3cm} cp{1.3cm} cp{1.3cm}}
     \hline
       & Stress & Work completion level & Average circle number & Average workload \\
     \toprule
     Competence mean & $-15.686^{***}$ & $0.740^{***}$ & $0.740^{***}$ & $0.064^{***}$ \\
     Competence difference & 0.144 & 0.028 & 0.031 & 0.003 \\
     N & 720 & 720 & 720 & 720 \\
     $R^{2}$ & 0.092 & 0.099 & 0.108 & 0.104 \\
     \bottomrule
    \end{tabular}
     \begin{tablenotes}
            \item Note:$^{*}p\textless0.1$,$^{**}p\textless0.05$,$^{***}p\textless0.001$
        \end{tablenotes}
    \label{tab:main_individial_statistics}
\end{table}

\subsection{Social network analysis}\label{social}

\begin{figure}
    \centering
    \includegraphics[width=0.8\linewidth]{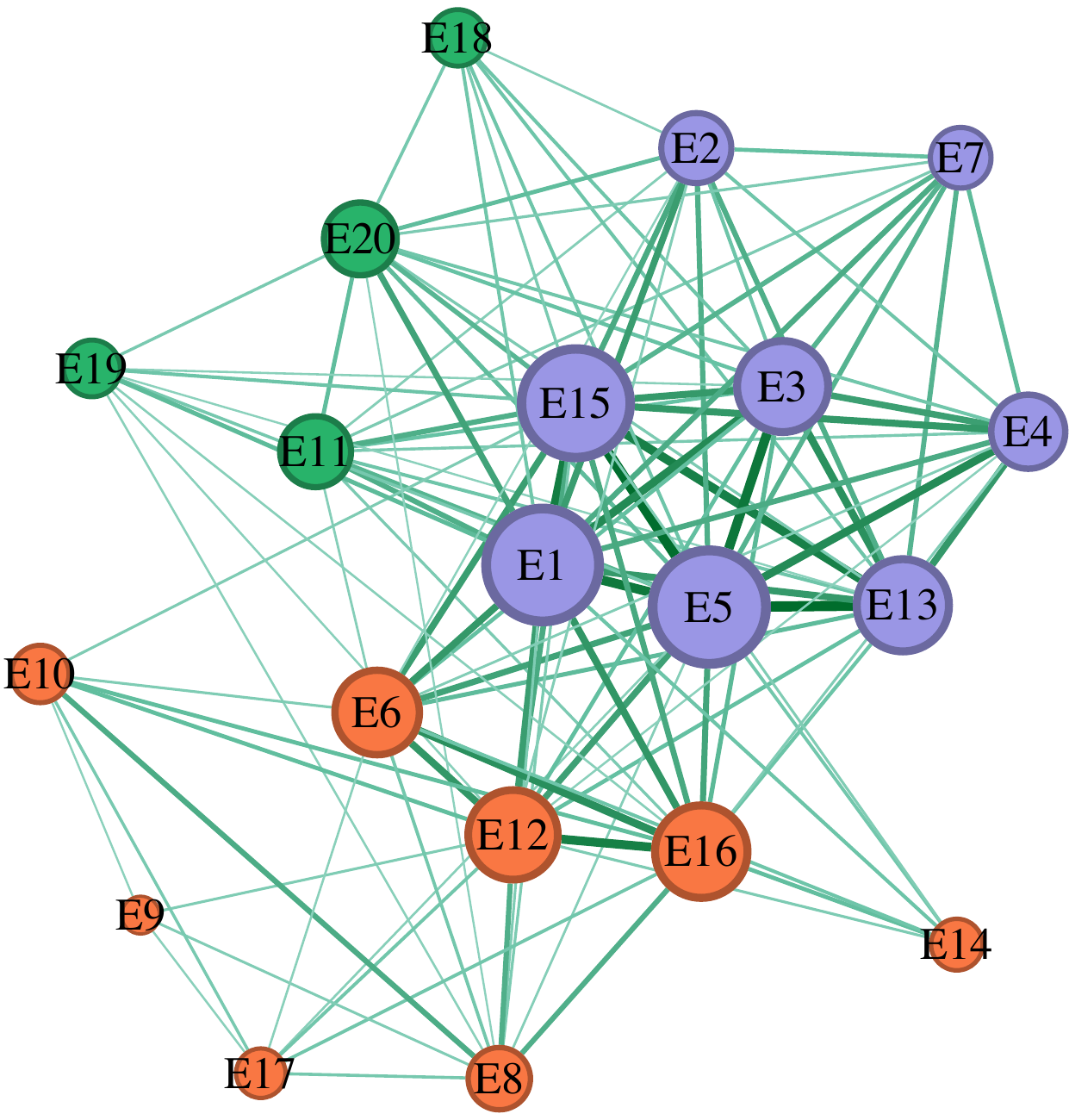}
    \caption{Social network condition at the end of the eighth week}
    \label{fig:social_network}
\end{figure}

To obtain detailed information about members' social behavior, we conducted a further analysis of social relationships within the organization. First, we constructed the organization's social network. We added an edge between members who participated in the same tasks over the eight weeks. To represent the strength of these connections in more detail, we used workload as the edge weight, indicating that fewer participants in a task can result in stronger connections between members. Ultimately, we obtained the organizational social network graph, as shown in Figure \ref{fig:social_network}. In the graph, we performed modularity-based clustering, with different colors representing different clusters, and the size of the nodes indicating their PageRank values. We can observe that over time, members form distinct sub-communities within the organization, with each sub-community having core members.

Furthermore, to examine whether members with higher status in the social network are related to their competences, we constructed an econometric model based on the individual-level analysis mentioned above (Equation \ref{eq:reg_individual}). The specific results are presented in Table \ref{tab:social_network_1} and Table \ref{tab:social_network_2}. According to the results, we found that both competences positively impact all social network indicators, including closeness centrality ($\beta_1=0.004, p<0.01; \beta_2=0.012, p<0.01$), betweenness centrality ($\beta_1=0.005, p<0.01; \beta_2=0.019, p<0.01$), eigenvector centrality ($\beta_1=0.013, p<0.01; \beta_2=0.039, p<0.01$), clustering coefficient ($\beta_1=0.014, p<0.01; \beta_2=0.044, p<0.01$), authority score ($\beta_1=0.003, p<0.01; \beta_2=0.009, p<0.01$), hub score ($\beta_1=0.003, p<0.01; \beta_2=0.009, p<0.01$) and PageRank value ($\beta_1=0.003, p<0.01; \beta_2=0.008, p<0.01$). Additionally, the positive impact of the competence mean on these indicators further proves the robustness of our results.
These findings indicate that individuals with stronger competence occupy more central positions in the social network. Furthermore, this result suggests that although the holacracy does not set fixed roles or include fixed task allocations, over time, more competent members tend to emerge and occupy important positions within the organization. This conclusion has significant implications for organizational structure design in management practice.
At the same time, differences in competences also have a significant impact on certain social network indicators. Competence differences have a significant positive effect on closeness centrality, betweenness centrality, and pagerank value. This indicates that having a higher competence in a particular aspect can also lead to members occupying central positions in the social network.


\begin{table}[h!]
    \centering
    \caption{Regression for social network at the 8th week at individual level}
    \begin{tabular}{p{1.3cm} p{1.3cm} p{1.3cm} p{1.3cm} p{1.3cm} p{1.3cm} p{1.3cm} p{1.3cm}}
     \hline
       & Closeness centrality & Betweenness centrality & Eigenvector centrality & Clustering & Authority score & Hub score & Pagerank\\
     \toprule
     Management competence & $0.004^{***}$ & $0.005^{*}$ & $0.013^{***}$ & $0.014^{***}$ & $0.003^{***}$ & $0.003^{***}$ & $0.003^{***}$ \\
     Functional competence & $0.012^{***}$ & $0.019^{***}$ & $0.039^{***}$ & $0.044^{***}$ & $0.009^{***}$ & $0.009^{***}$ & $0.008^{***}$ \\
     N & 720 & 720 & 720 & 720 & 720 & 720 & 720 \\
     $R^{2}$ & 0.219 & 0.073 & 0.259 & 0.172 & 0.262 & 0.262 & 0.262 \\
     \bottomrule
    \end{tabular}
     \begin{tablenotes}
            \item Note:$^{*}p\textless0.1$,$^{**}p\textless0.05$,$^{***}p\textless0.001$
        \end{tablenotes}
    \label{tab:social_network_1}
\end{table}

\begin{table}[h!]
    \centering
    \caption{Regression for social network at the 8th week at individual level of competence statistics}
    \begin{tabular}{p{1.3cm} p{1.3cm} p{1.3cm} p{1.3cm} p{1.3cm} p{1.3cm} p{1.3cm} p{1.3cm}}
     \hline
       & Closeness centrality & Betweenness centrality & Eigenvector centrality & Clustering & Authority score & Hub score & Pagerank\\
     \toprule
     Competence mean & $0.016^{***}$ & $0.023^{***}$ & $0.052^{***}$ & $0.058^{***}$ & $0.012^{***}$ & $0.012^{***}$ & $0.010^{***}$ \\
     Competence difference & $0.001^{**}$ & $0.004^{***}$ & 0.001 & -0.001 & 0.000 & 0.000 & $0.000^{**}$ \\
     N & 720 & 720 & 720 & 720 & 720 & 720 & 720 \\
     $R^{2}$ & 0.181 & 0.073 & 0.210 & 0.135 & 0.212 & 0.212 & 0.214 \\
     \bottomrule
    \end{tabular}
     \begin{tablenotes}
            \item Note:$^{*}p\textless0.1$,$^{**}p\textless0.05$,$^{***}p\textless0.001$
        \end{tablenotes}
    \label{tab:social_network_2}
\end{table}
\section{Conclusion}\label{sec4}
This paper develops an LLM-based simulation framework to address structural issues in organizational science research that are difficult to explore experimentally. The framework leverages the powerful cognitive and behavioral capabilities of LLMs. The analysis of the experimental results leads to three main conclusions:
First, at the organizational level, an increase in average competence reduces average stress level but has limited impact on enhancing the competence coded in the organization. Specifically, the average task completion level within the organization decreases.
Second, at the individual level, an increase in competence also reduces individual stress. Unlike organizations, individuals' task completion levels improve with higher competence. Further analysis of social structures reveals that individuals with higher competence tend to be self-driven and more goal-oriented in their task selection.
Third, in terms of social networks, individuals with higher competence occupy more central positions. Higher competence in a specific area also leads to similar effects.
Theoretically, these conclusions contribute to research on organizational structures in organizational science, particularly studies on decentralized organizations. Our framework provides an experimental foundation for research on holacracy, allowing for various parameter adjustments to conduct further experiments related to holacracy. Practically, our experimental results offer managerial insights for the operation of holacracy and provide guidance on how managers can adjust organizational structures. Additionally, the dynamic analysis of the organizational social network enhances our understanding of organizational dynamics.

%





\bibliography{reference}

\end{document}